# Thermal-induced ion magnetic moment in $H_4O$ superionic state

Xiao Liang[1], Junhao Peng[1], Fugen Wu[2*], Renhai Wang[1], Yujue Yang[1], Xingyun Li[3], Huafeng Dong[1*]

[1] Guangdong Provincial Key Laboratory of Sensing Physics and System Integration Applications, School of Physics and Optoelectronic Engineering, Guangdong University of Technology, Guangzhou 510006, China

[2] The College of Information Engineering, Guangzhou Vocational University of Science and Technology, Guangzhou 510550, China

[3] DongGuan Institute of GuangDong Institute of Mertology, Dongguan 523343, China

The hydrogen ions in the superionic ice can move freely, playing the role of electrons in metals. Its electromagnetic behavior is the key to explaining the anomalous magnetic fields of Uranus and Neptune. Based on the *ab initio* evolutionary algorithm, we searched for the stable $H_4O$ crystal structure under pressures of 500-5000 GPa and discovered a new layered chain *Pmn*$2_1$-$H_4O$ structure with $H_3$ ion clusters. Interestingly, $H_3$ ion clusters rotate above 900 K (with an instantaneous speed of 3000 m/s at 900 K), generating an instantaneous magnetic moment ($\sim 10^{-26}$ A·m$^2 \approx 0.001$ $\mu_B$). Moreover, H ions diffuse in a direction perpendicular to the H-O atomic layer at 960-1000 K. This is because the hydrogen oxygen covalent bonds within the hydrogen oxygen plane hinder the diffusion behavior of $H_3$ ion clusters within the plane, resulting in the diffusion of $H_3$ ion clusters between the hydrogen oxygen planes and the formation of a one-dimensional conductive superionic state. One-dimensional diffusion of ions may generate magnetic fields. We refer to these two types of magnetic moments as "thermal-induced ion magnetic moments". When the temperature exceeds 1000 K, H ions diffuse in three directions. When the temperature exceeds 6900 K, oxygen atoms diffuse and the system becomes fluid. These findings provide important references for people to re-recognize the physical and chemical properties of hydrogen and oxygen under high pressure, as well as the sources of abnormal magnetic fields in Uranus and Neptune.

Introduction

Uranus and Neptune have special non-polar and non-axisymmetric magnetic fields, and the source of this anomalous magnetic field is reported to come from the middle ice layers of Uranus and Neptune[1, 2]. Exploring the structure and electrical properties of the middle ice layers is a key challenge in explaining the source of anomalous magnetic fields. In order to explore the structure and electrical properties of the middle ice layers, researchers have studied the structure and electrical properties of ice at different temperatures and pressures[3-11].

In 1999, Cavazzoni *et al*. reported that in the middle layer planetary ices are in the fluid state: (i) a molecular regime at low pressure, (ii) a nonmetallic ionic regime at intermediate pressure and temperature, and (iii) a metallic regime at high pressure and temperature[11]. In 2013, Wilson *et al*. discovered a new superionic ice structure (fcc) with almost zero energy difference between bcc and fcc at 1 Mbar and 3000 K. At higher pressures, the stability of fcc structure is stronger than that of bcc structure[4]. The middle layer of Uranus and Neptune is mainly composed of $H_2O$, $CH_4$, and $NH_3$[12]. Under high pressure, $CH_4$ decomposes to produce $H_2$, which provides conditions for the formation of hydrogen rich HO compounds. Therefore, Ma *et al*. studied the structure and superionic state of $H_3O$ under the pressure and temperature environment inside Uranus and Neptune, and explored the influence of hydrogen rich H-O compounds on the magnetic field[13]. In experiments, Vos *et al*. revealed the formation of a novel hydrate $H_4O$ at pressures ranging from 2.3-30 GPa through high-pressure optical and X-ray studies [14]. In theory, the miscibility of $H_2$ and $H_2O$ has also been studied at 2-70 GPa and 1000-6000 K [15]. Although numerous studies have reported the structure of H-O system under high pressure and the superionic ice states, the exploration of the stable structure of $H_4O$ is still insufficient, especially the relationship between the transition of superionic state and the magnetic field is still unclear.

In this article, we report a new $Pmn2_1$-$H_4O$ phase and investigate its special dynamic behavior in the superionic state at high temperature and high pressure. We found that the H ions in the $H_3$ ion clusters in the $H_4O$ superionic state undergoes rotation and one-dimensional diffusion, which may generate a magnetic field due to the movement of charged ions, which we call the "thermal-induced ion magnetic moment". This discovery will provide important references for people to re-recognize the physical and chemical properties of hydrogen and oxygen under high pressure, as well as the sources of abnormal magnetic fields in Uranus and Neptune.

Calculation method

We used *ab*-initio quantum-mechanical calculations USPEX to search for $H_4O$ structures under pressures of 500-4000 GPa[16]. For each pressure point, the first generation of 150 structures were produced by random symmetric structure generator. 50 structures were calculated for each subsequent generation, which were generated through heredity (20%), random symmetric structure generator (20%), soft mutation or coomutation (20%), transmutation (20%), lattice mutations (10%), and topological random generator (10%) methods. Structural optimization and energy calculation were performed using the density functional theory based software VASP [17]. Calculations were performed in the GGA approximation. Projector augmented wave (PAW) pseudopotential and the Perdew-Burke-Ernzerhof exchange-correlation functional are used. The pseudopotential cutoff radii are 0.8 and 1.1 Bohr for H and O. The Brillouin zone sampling adopts a density automatic uniform distribution of 0.251. The cut-off energy for structure search is set to 500 eV, while the cut-off energy for subsequent calculations is set to 1700 eV, and the total energy converges to 1 meV.

We used ab initio molecular mechanics to simulate the dynamic behavior at high temperatures. The temperature range for molecular dynamics simulation is 300-8000 K. The calculated $Pmn2_1$-$H_4O$ unit cell size is: *a*=8.2260 Å, *b*=7.3158 Å, *c*=8.6568 Å, containing 480 atoms. *k*-point sampling is performed with the Gamma point as the center. NVT simulation used Nose Hover thermostat[18-20]. Each temperature is calculated with a time step of 0.1 fs for 100000 steps or 150000 steps.

Results and Discussion

Figure 1 shows the convex hull diagram of hydroxide compounds containing $H_2$, $H_4O$, and $H_2O$, as well as the phase diagram of the enthalpy value of $H_4O$ structure as a function of pressure. The results show that when the pressure is above 1440 GPa, the structure of $H_4O$ can exist stably without decomposing into $H_2$ and $H_2O$, which is consistent with the prediction of Zhang *et al.*[21]. Interestingly, a new and stable $Pmn2_1$-$H_4O$ structure has emerged within the pressure range of 1440-1570 GPa (see Table SI for the structural parameters of $Pmn2_1$ and $Pnma$ structures). When the pressure exceeds 1570 GPa, $Pmn2_1$ structure transforms into $Pnma$ structure. These two structures are both indirect bandgap semiconductors in the pressure range of 1440-1570 GPa. At 1500 GPa, the bandgap widths of $Pmn2_1$ and $Pnma$ structures are 2.85 eV and 2.88 eV, respectively, and decrease with increasing pressure (FIGS. S1 and S2).

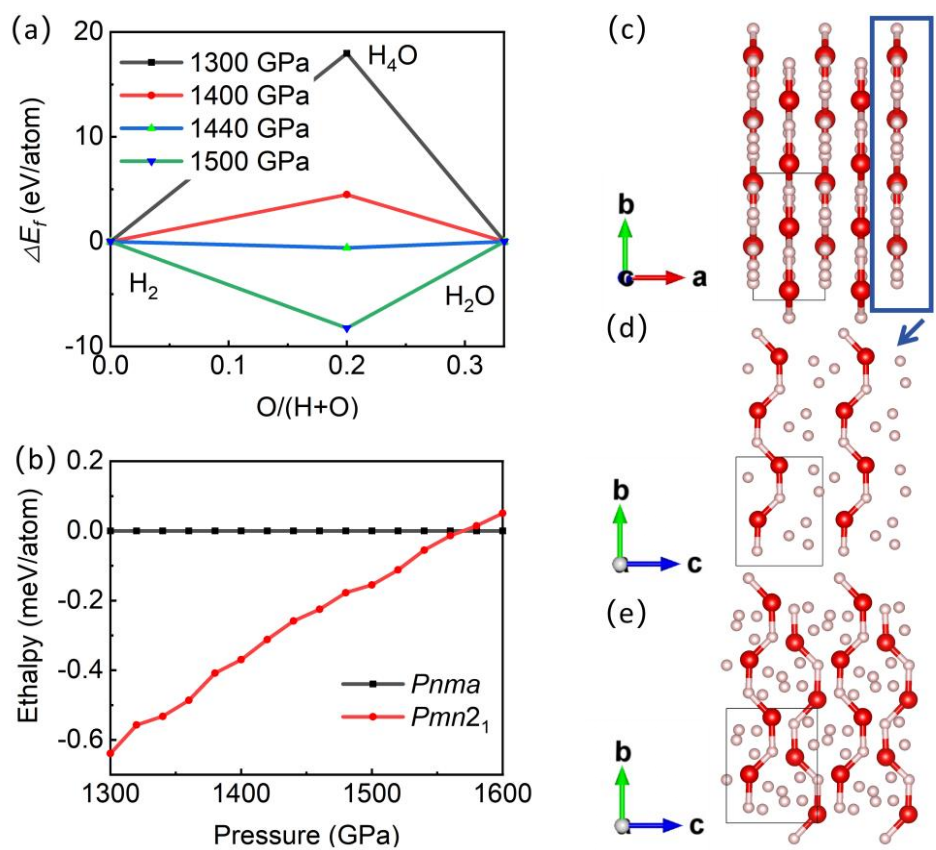

FIG.1 Phase diagram of hydrogen-rich H-O compounds and crystal structure of new structure. (a)

Enthalpy of hydrogen-rich H-O compounds at different pressures, relative to a mixture of $H_2$ and $H_2O$. (b) Enthalpy versus pressure of different $H_4O$ structures. (c-e) Structure of $Pmn2_1$ under 1500 GPa. The red sphere in the (c-e) represents oxygen atoms, and the pink sphere represents hydrogen atoms.

The new $H_4O$ phase is an orthorhombic system, AB stacking layered structure (FIG. 1c), with $Pmn2_1$ space group ($a$=2.056 Å, $b$=3.658 Å, $c$=2.886 Å at 1500 GPa). Each layer has $H_3$ ion clusters connected by H-H covalent bonds (bond lengths are 0.707 Å, 0.724 Å, and 0.741 Å, respectively) to form triangles. $H_3$ ion clusters are distributed between parallel -H-O-H-O- chains (bond lengths of H-O are 1.098 Å, 1.027 Å, 1.056 Å, and 1.070 Å, respectively) (FIG. 1d). The -H-O-H-O- chains of adjacent layers are arranged in reverse (FIG. 1e). The distance between atomic layers is 1.028 Å.

By analyzing Electron Localization Function (ELF), we can understand the bonding and interactions of new structures, with high ELF values indicating strong bonding[22, 23]. The ELF of the (100) crystal plane of $Pmn2_1$-$H_4O$ (FIG. 2a) shows that a high electron localization is observed in the $H_3$ ionic cluster to form H-H covalent bonds. The low density of electrons distributed between the $H_3$ ion cluster and the adjacent oxygen atoms indicates that the $H_3$ ion cluster is relatively independent and does not form strong chemical bonds with surrounding atoms, which provide possibilities for appearing the rotation phenomenon of $H_3$ ion clusters caused by temperature rise. There is a shared electron between the hydrogen and oxygen atoms on the -H-O-H-O- chain, forming a strong H-O covalent bond. In the Electron Localization Function of the (001) crystal plane (FIG. 2b), the charge density between H-O atomic layers is low, and the interaction between atomic layers is weak. The atomic layers are relatively independent.

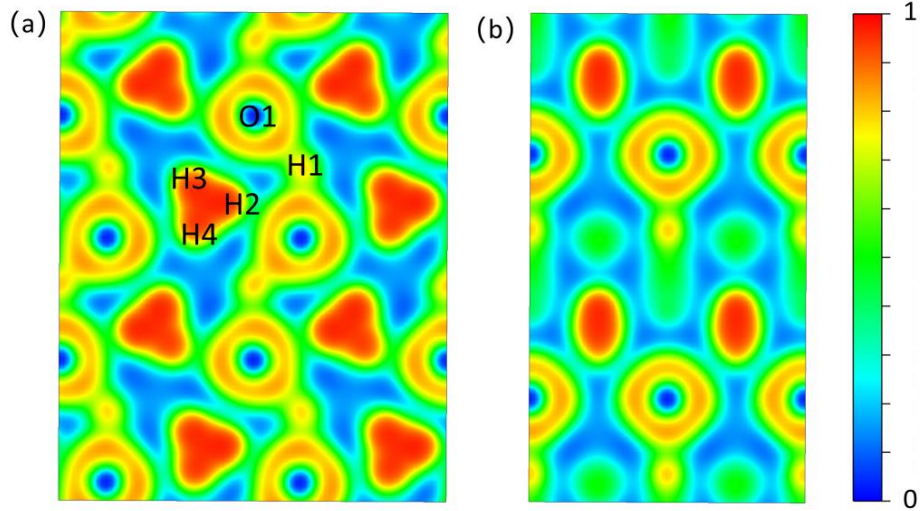

FIG. 2. The ELF isosurfaces of *Pmn*2$_1$. (a) (1 0 0) crystal face. (b) (0 0 1) crystal face.

In order to analyze the stability of *Pmn*2$_1$ structure at different temperatures and the diffusion of hydrogen and oxygen atoms, we conducted molecular dynamics simulations on *Pnm*2$_1$ structure at different temperatures under a pressure of 1500 GPa. The curve of atomic mean square displacement and total energy over time at room temperature (300 K) (FIG. S3) shows that the mean square displacement of oxygen atoms is close to 0, and the mean square displacement of hydrogen atoms is about 0.25 Å$^2$, indicating that both oxygen and hydrogen atoms vibrate near the atomic position without damaging the stability of the structure, and the system is stable in the solid(diffusion coefficients $D_O=0$, $D_H=0$). When the temperature rises to 800 K (FIG. 3a), the mean square displacement of both oxygen and hydrogen atoms increases, indicating that the vibrations of both oxygen and hydrogen atoms have increased, but the structure remains stable (diffusion coefficients $D_O=0$, $D_H=0$).

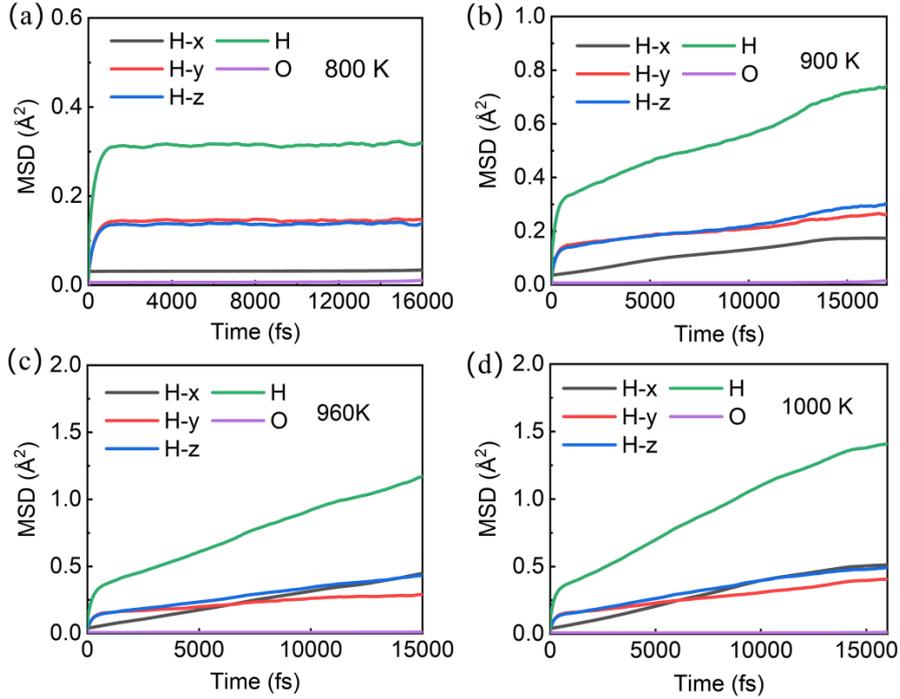

FIG. 3. Mean Square Displacement of Atoms in $Pmn2_1$-$H_4O$ at Different Temperatures. (a)800 K. (b) 900 K. (c) 960 K. (d) 1000 K.

When the temperature reaches 900 K, the mean square displacement of hydrogen atoms increases to ~0.74 Å$^2$ (FIG. 3b), and remained at ~0.17 Å$^2$ in the X direction (diffusion coefficient $D_{Hx}$=0), indicating that hydrogen atoms did not diffuse in the X direction (the trajectory diagram of H ions in FIG. 4a confirm that H ions do not diffuse in the X direction). However, the mean square displacement in the Y and Z directions is ~0.26 Å$^2$ and 0.30 Å$^2$, respectively. Therefore, the root mean square displacement of hydrogen atoms in the YZ plane is ~0.75 Å, which is greater than the bond length of H-H in the $H_3$ ion cluster, indicating that the H ions in the $H_3$ ion cluster may move among the three H ions positions. In order to identify whether it is move or not, we calculate the trajectory diagram of H ions at 900 K (FIG. 4b). The trajectory diagram of the H ions in the $H_3$ ion cluster is a circular curve, indicating that the $H_3$ ion cluster has rotated, which confirmed by the atomic state diagram at 1.22-1.25 ps (Figure S4). The instantaneous velocity of H ions rotation is about 3000 m/s, as it is known from the Bader charge analysis that each H in the $H_3$ ion cluster carries a charge of ~ +0.22e (Table SII). In addition, due to the magnitude of the magnetic moment generated by the rotational motion of ions, $\mu = IS = \frac{qvr}{2}$, where $q$ is the charge of the ion, $v$ is the velocity of the charged ion, and $r$ is the orbital radius of the charged ion's rotation. The estimated instantaneous magnetic moment generated by the rotation of a single $H_3$ ion cluster is approximately $10^{-26}$ A·m$^2 \approx 0.001$ $\mu_B$, which we refer to as the "thermal-induced ion magnetic moment". Because the rotation time and direction of the $H_3$ ion cluster are randomly generated, the superionic state as a whole does not exhibit magnetism. But if we can control the rotation direction of $H_3$ ion clusters, the magnetism of superionic state can be controlled. For the atoms in the -H-O-H-O- chain, it can be inferred from FIG. 3b and FIG. 4ab that they vibrate near their equilibrium positions.

When the temperature reaches 960 K, the mean square displacement of hydrogen atoms increases to about 1.17 Å$^2$ at 15 ps, and the direction with the largest increase in mean square displacement is in the X direction (FIG. 3c), suggesting that hydrogen atoms may diffuse in this direction. Through the trajectory diagram of H ions (FIG. 4cd), it can be determined that the H ions in the $Pmn2_1$ structure diffuse only in the X direction. This diffusion phenomenon originates from the movement of H ions between $H_3$ ion clusters in different layers, while the H ions on the -H-O-H-O- chain do not diffuse due to the constraint of hydrogen oxygen covalent bonds. At the same time, the -H-O-H-O- chain also constrains the diffusion of $H_3$ ion clusters in the plane. At this point, with $D_O=0$ and $D_H=0.9\times10^{-10}$ m$^2$·s$^{-1}$, the O atom remains near the equilibrium position, forming a sublattice. The H ions begin to diffuse, and the system transitions to a superionic state ($D_O=0$, $D_H>0$). This is a special superionic state, where H ions can only move in the X direction, that is, this is a one-dimensional conductive superionic state.

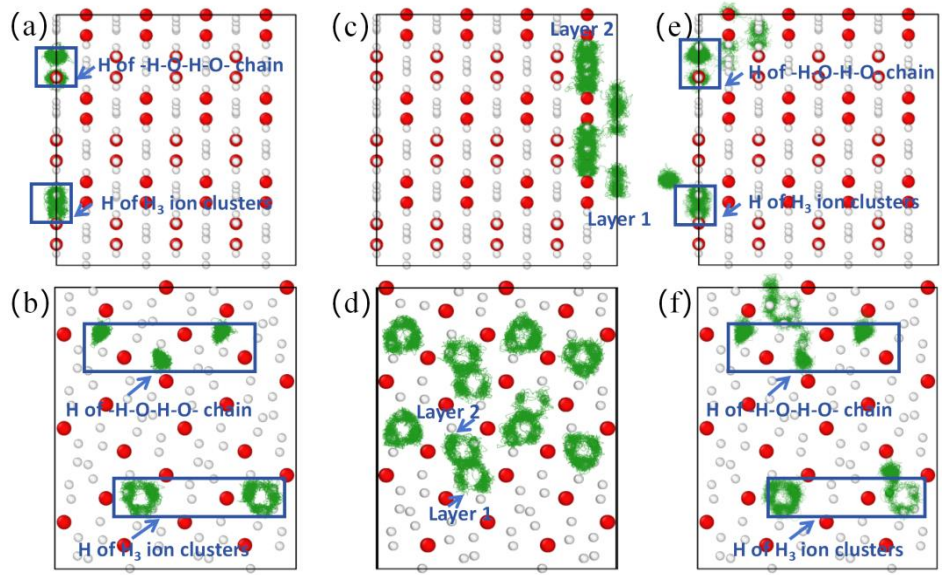

FIG. 4. Atomic trajectory diagrams of (0 1 0) and (1 0 0) crystal face of $Pmn2_1$-$H_4O$ at different temperatures. (a, b) Trajectory diagram of H ions at 900 K. (c, d) Trajectory diagram of H ions in $H_3$ ion cluster at 960 K. (e, f) Trajectory diagram of H ions at 1000 K.

When the temperature reaches 1000 K, the mean square displacement of O atoms remains zero, and H ions is ~1.5 Å$^2$ at 15 ps (FIG. 3d), with $D_H=1.1\times10^{-10}$ m$^2\cdot$s$^{-1}$. The H ions in the -H-O-H-O- chain can move to the position of the neighboring $H_3$ ion clusters (FIG. 4ef), because as the temperature increases, the atomic kinetic energy increases, and the energy is greater than the covalent bond energy of H-O, thereby breaking the -H-O-H-O- chain. At this point, the O atom remains near the equilibrium position to form a sub lattice, while the H ions can diffuse in three directions. It is worth noting that the diffusion velocity of H ions exhibits significant anisotropy in three directions, with the fastest diffusion occurring in the X direction perpendicular to the H-O atomic layer.

When the temperature reaches 6900 K, $D_O>0$, $D_H>0$, Both H and O atoms diffuse, and the system becomes fluid. And as the temperature increases, the mean square displacement and diffusion coefficient of O atoms rapidly increase (FIG. S5).

Conclusion

We have discovered a new layered chain $Pmn2_1$-$H_4O$ structure, which is more stable than the reported *Pnma* structure at 1440-1570 GPa. At the same time, the structure maintains dynamic stability at temperatures below 900 K. When the temperature is above 900 K, $H_3$ ion clusters rotate in the hydrogen oxygen plane at an instantaneous speed of 3000 m/s, and the generated instantaneous magnetic moment is about $10^{-26}$ A·m$^2 \approx 0.001$ $\mu_B$. When the temperature is 960 K, the H ions in the $H_3$ ion cluster diffuse between the layers of hydrogen and oxygen atoms, resulting in the H ions diffusing only in one direction, leading to a special superionic state of one-dimensional conductivity. When the temperature exceeds 1000 K, the hydrogen oxygen covalent bond on the -H-O-H-O- chain is broken, and the H ions can diffuse in three directions. When the temperature is above 6900 K, the diffusion coefficient of oxygen atoms is greater than 0, and hydrogen and oxygen atoms can freely diffuse in three directions to form a fluid. This discovery will provide important references for people to re-recognize the physical and chemical properties of hydrogen and oxygen under high pressure, as well as the sources of abnormal magnetic fields in Uranus and Neptune.

Acknowledgements

This work was supported by the Guangdong Natural Science Foundation of China (Grant No. 2017B030306003), the National Natural Science Foundation of China (Grant No. 11604056) and the State Administration for Market Regulation Science and technology project (Grant No. 2023MK093). The authors thank the Center of Campus Network & Modern Educational Technology, Guangdong University of Technology, Guangdong, China, for providing computational resources and technical support for this work.

Author Contributions

Conceptualization, Huafeng Dong and Xiao Liang; methodology, Huafeng Dong and Xiao Liang; data curation, Xiaoliang, Junhao Peng and Xingyu Li; supervision, Fugen Wu and Renhai Wang and Yujue Yang and Huafeng Dong; writing—original